\documentclass[aps,prb,twocolumn,floatfix,citeautoscript,superscriptaddress]{revtex4}
\usepackage{amsmath}
\usepackage{graphicx}
\usepackage{dcolumn}
\usepackage{float}
\usepackage{color}
\usepackage{multirow}
\usepackage{amssymb}
\usepackage{microtype}
\usepackage{booktabs}
\usepackage{hyperref}

\usepackage[a4paper,margin=0.9in]{geometry}

\begin{document}

\title{Two types of charge order in the superconducting kagome material CsV$_3$Sb$_5$}

\author{Ritu Gupta\footnotemark[4]}
\email{ritu.gupta@psi.ch}
\thanks{Equal contribution}
 
  \affiliation{Laboratory for Muon Spin Spectroscopy, Paul Scherrer Institute, CH-5232 Villigen PSI, Switzerland}

 \author{Debarchan Das\footnotemark[4]}
 \email{debarchandas.phy@gmail.com}
\thanks{Equal contribution}
 \affiliation{Laboratory for Muon Spin Spectroscopy, Paul Scherrer Institute, CH-5232 Villigen PSI, Switzerland}
 
 \author{Charles Mielke III}
 \affiliation{Laboratory for Muon Spin Spectroscopy, Paul Scherrer Institute, CH-5232 Villigen PSI, Switzerland}
 
  \author{Ethan Ritz}
 \affiliation{Department of Chemical Engineering and Materials Science, University of Minnesota, MN 55455, USA}
 
  \author{Fabian Hotz}
 \affiliation{Laboratory for Muon Spin Spectroscopy, Paul Scherrer Institute, CH-5232 Villigen PSI, Switzerland}
 
\author{Qiangwei Yin}
\affiliation{Department of Physics and Beijing Key Laboratory of Opto-electronic Functional Materials \& Micro-nano Devices, Renmin University of China, Beijing 100872, China}
\author{Zhijun Tu}
\affiliation{Department of Physics and Beijing Key Laboratory of Opto-electronic Functional Materials \& Micro-nano Devices, Renmin University of China, Beijing 100872, China}
\author{Chunsheng Gong}
\affiliation{Department of Physics and Beijing Key Laboratory of Opto-electronic Functional Materials \& Micro-nano Devices, Renmin University of China, Beijing 100872, China}
\author{Hechang Lei}
\email{hlei@ruc.edu.cn}
\affiliation{Department of Physics and Beijing Key Laboratory of Opto-electronic Functional Materials \& Micro-nano Devices, Renmin University of China, Beijing 100872, China}
 \author{Turan Birol}
 \affiliation{Department of Chemical Engineering and Materials Science, University of Minnesota, MN 55455, USA}
  \author{Rafael M. Fernandes}
 \affiliation{School of Physics and Astronomy, University of Minnesota, Minneapolis, MN 55455, USA}

   \author{Zurab Guguchia}
 \affiliation{Laboratory for Muon Spin Spectroscopy, Paul Scherrer Institute, CH-5232 Villigen PSI, Switzerland} 
    \author{Hubertus Luetkens}
    \email{hubertus.luetkens@psi.ch}
 \affiliation{Laboratory for Muon Spin Spectroscopy, Paul Scherrer Institute, CH-5232 Villigen PSI, Switzerland}
\author{Rustem Khasanov}
 \email{rustem.khasanov@psi.ch}
 \affiliation{Laboratory for Muon Spin Spectroscopy, Paul Scherrer Institute, CH-5232 Villigen PSI, Switzerland}

\date{\today}

\maketitle

\textbf{The kagome metals of the family $A$V$_3$Sb$_5$, featuring a unique structural motif, harbor an array of intriguing phenomena such as chiral charge order and superconductivity. CsV$_3$Sb$_5$ is of particular interest because it displays a double superconducting dome
in the region of the temperature-pressure phase diagram where charge order is still present. However, the microscopic origin of such an unusual behavior remains an unsolved issue. Here, to address it, we combine high-pressure, low-temperature muon spin relaxation with first-principles calculations. We observe a pressure-induced threefold enhancement of the superfluid density, which also displays a double peak feature, similar to the superconducting critical temperature. This leads to three distinct regions in the phase diagram, each of which features distinct slopes of the linear relation
between superfluid density and the critical temperature. These results are attributed to a possible evolution of the charge order pattern from the superimposed tri-hexagonal Star-of-David phase at low pressures (within the first dome) to the staggered tri-hexagonal phase at intermediate pressures (between the first and second domes). Our findings suggest a change in the nature of the charge ordered state across the phase diagram of CsV$_3$Sb$_5$, with varying degrees of competition with superconductivity.}

 Among the series $A$V$_3$Sb$_5$ ($A$ = Rb, K, Cs) \cite{Neupert2022review,Jiang2021review,mielke2021time}, the Cs compound manifests the highest superconducting critical temperature $T_c$ $\simeq$ 2.5 K. CsV$_3$Sb$_5$ also features multi-gap superconductivity and more importantly, a time reversal symmetry breaking (TRSB) chiral charge order below $T_{co}$ = 94~K \cite{yu2021evidence,Wukerr,RustemCVS,wang2021electronic}, as reported by scanning tunneling microscopy \cite{zhao2021}, polar Kerr rotation \cite{Wukerr}, and $\mu$SR experiments\cite{yu2021evidence,RustemCVS}. A comprehensive understanding of the interdependence between charge order (CO) and superconductivity (SC) is thus essential, and can be studied by using an appropriate external perturbation. In the quest to obtain an efficient tuning knob, hydrostatic pressure was found to be optimal. Indeed, pressure suppresses the charge order and results into an unusual but well-pronounced double superconducting dome in the temperature-pressure phase diagram. Compared to the Rb and K counterparts \cite{wang2021competition,zhu2021double,du2021pressure}, the double peak behavior is most distinguishable in the case of CsV$_3$Sb$_5$ \cite{yu2021unusual,wang2021charge,chen2021double,zhang2021pressure} where $T_c$ is roughly tripled to 8 K around the optimal pressure of 2 GPa from its value $T_c$ = 2.5~K at ambient pressure. Thus, pressure tuned CsV$_3$Sb$_5$ offers a rich framework for studying the nature of the interplay between superconductivity and charge order.

 To systematically characterize and to obtain a microscopic understanding of the complex temperature-pressure phase diagram of CsV$_3$Sb$_5$, here we report high-pressure zero-field and transverse-field ${\mu}$SR as well as AC susceptibility (up to 1.75 GPa) measurements combined with first principles calculations. Transverse field (TF) ${\mu}$SR experiments serve as an extremely sensitive local probe technique to measure the magnetic penetration depth $\lambda$ in the vortex state of bulk type II superconductors. This quantity is directly related to the superfluid density $n_{s}$ via 1/${\lambda}^{2}$~=~$\mu_{0}$$e^{2}$$n_{s}/m^{*}$ (where $m^{*}$ is the effective mass). Zero-field ${\mu}$SR has the ability to detect internal magnetic fields as small as 0.1 G without applying external magnetic fields, making it a highly valuable tool for probing spontaneous magnetic fields due to TRSB in exotic superconductors. The techniques of ${\mu}$SR and DFT complement each other ideally, as we are able to sensitively probe the fundamental microscopic properties of CsV$_3$Sb$_5$ with ${\mu}$SR experiments and correlate them with the pressure-evolution of the charge ordered state calculated by DFT.

\begin{figure*}[htb!]
\includegraphics[width=1\linewidth]{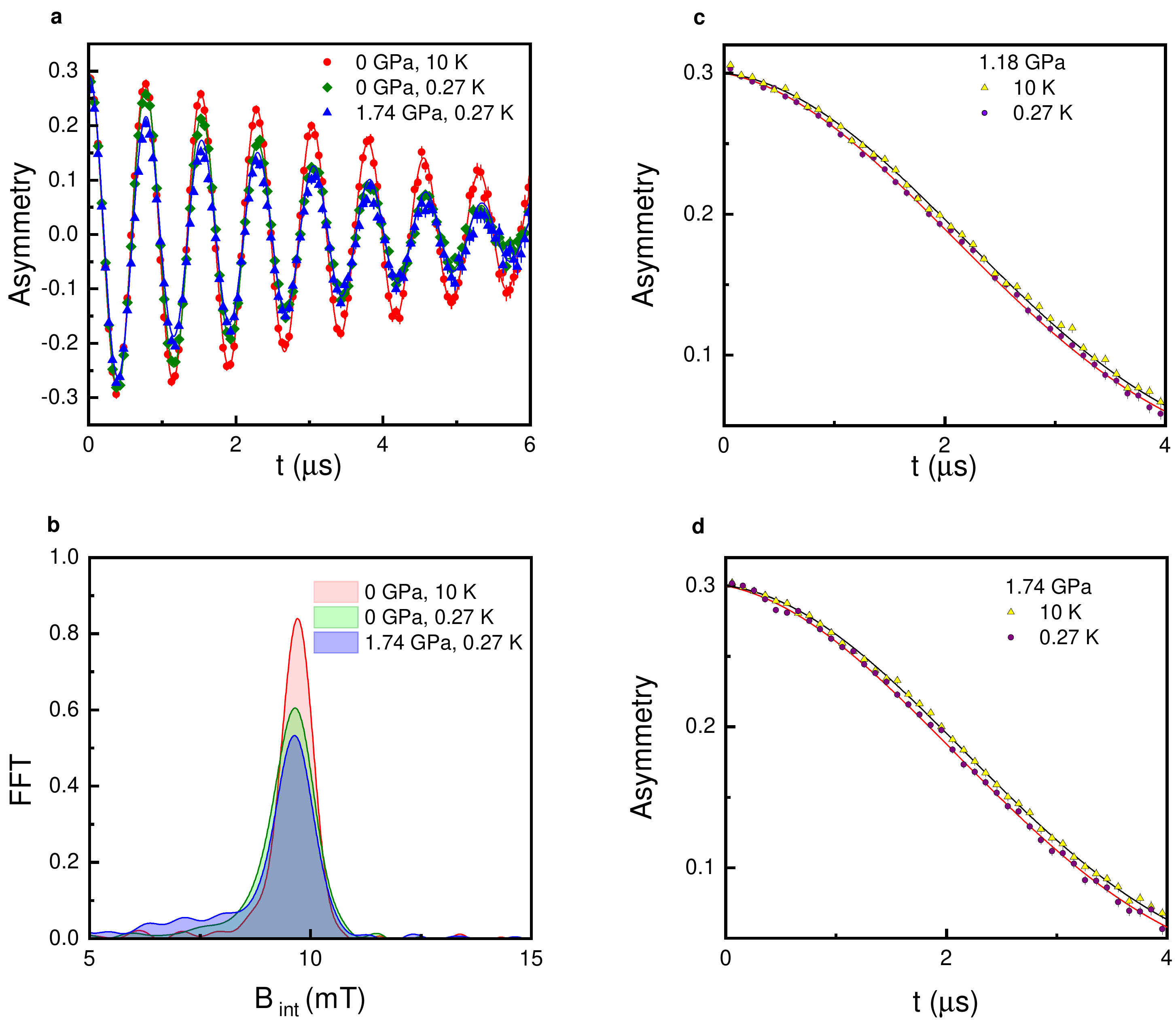}
\caption{(Color online) {\bf TF and ZF-$\mu$SR time spectra.} a.) Selected TF-$\mu$SR time spectra at 0.27~K (in the superconducting state) and 10~K (in the normal state) collected for ambient and 1.74~GPa pressure. Solid lines are the fits of the spectra considering both the sample and background parts, as described in the Supplementary section. b.) The fast Fourier transform of the spectra shows an additional broadening of the peak due to an inhomogeneous distribution of internal fields, resulting from the appearance of a vortex lattice. c.)-d.) Representative ZF-$\mu$SR time spectra collected in zero field well below $T_c$ (0.27~K) and above $T_c$ (10~K) for 1.18~GPa [panel (c)] and 1.74~GPa [panel (d)]. The solid lines are fits considering two relaxation channels, namely Gaussian-Kubo-Toyabe and a Lorentzian decay function.  }
\label{fig:1}
\end{figure*}
\begin{figure*}[htb!]
\includegraphics[width=1\linewidth]{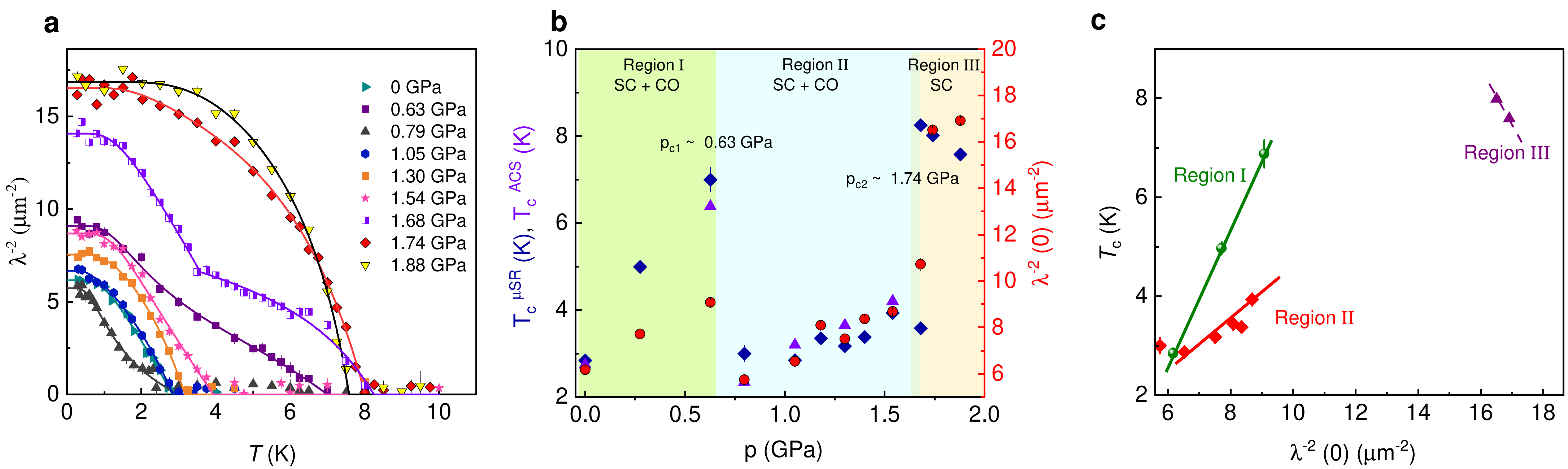}
\caption{(Color online) {\bf Pressure dependence of the superconducting parameters and temperature-pressure phase diagram.} a.) Temperature variation of the inverse squared magnetic penetration depth $\lambda^{-2}(T)$ for the applied hydrostatic pressures shown in the plot legend. The solid line is a two gap $s+s$-wave model fitting using the phenomenological $\alpha$-model. 
b.) The temperature-pressure superconducting phase diagram of CsV$_3$Sb$_5$. The left axis shows $T_c$ values obtained through TF-$\mu$SR and ac magnetic susceptibility measurements carried out under hydrostatic pressure conditions. The right axis presents the inverse squared London penetration depth at base temperature, i.e., $\lambda^{-2}(0)\propto n_s$ as a function of pressure. c.) Superconducting transition temperature $T_c$ plotted as a function of the base-temperature superfluid density $\lambda^{-2}(0)$ for the three different regions of the phase diagram of panel (b).}
\label{fig:2}
\end{figure*}
\section{results}
\subsection{Superfluid density as a function of pressure}

Figure 1(a) shows the TF-$\mu$SR time spectra of  CsV$_3$Sb$_5$ well below (0.25~K) and above (10~K) $T_c$ in the presence of an applied magnetic field of 10~mT  for ambient as well as the highest applied pressure, 1.74~GPa. The weakly damped oscillations above $T_c$ are essentially due to the randomly distributed local fields created by nuclear magnetic moments. The spectra below $T_c$ show a sizable increase in relaxation because of the inhomogeneous distribution of internal fields arising due to the formation of the flux line lattice in the vortex state. The Fourier transform of the spectra, shown in Fig. 1b, gives the internal field distribution, which showcases a clear broadening in the superconducting state compared to the normal state. 
From the TF-$\mu$SR data, the superconducting relaxation rate $\sigma_{\rm sc}(T)$ is extracted. The latter is related to the magnetic penetration depth via
$\frac{\sigma_{\rm sc}(T)}{\gamma_\mu}=0.06091\frac{\phi_0}{\lambda^2(T)}$, where $\gamma_\mu$~=~2$\pi\times$135.5~MHz/T is the muon gyromagnetic ratio and $\phi_0$ is the flux quanta \cite{brandt1988flux,brandt2003properties} (details of the analysis of the TF-$\mu$SR data are given in the Supplementary section). The resulting temperature dependence of the inverse squared magnetic penetration depth $\lambda^{-2}(T)$, measured under various hydrostatic pressures, are shown in Fig. 2(a).

 The $\lambda^{-2}(T)$ curves in Fig. 2(a) are well described by a two-gap model across the entire pressure range probed (see details in the Methods section). 
This indicates that the multi-gap nature of the superconducting state of CsV$_3$Sb$_5$, which was previously reported by various techniques for ambient conditions \cite{xu2021multiband,duan2021nodeless,gupta2021microscopic,nakayama2021multiple}, is robust against the application of hydrostatic pressure. As we show below, a sharp increase of both $T_c$ and $\lambda^{-2}(0)$ occurs at pressures in the range 1.5~--~1.7~GPa, suggesting the complete suppression of the CO and thus a transition between the SC+CO state to the pure SC state. It is quite remarkable, therefore, that at the intermediate pressure 1.68~GPa, which is at the border between the SC+CO and pure SC phases, the temperature evolution of the penetration depth shows a prominent two-step like feature [purple curve in Fig.~2(a)]. This is a signature of inhomogeneity, whose origin can be either extrinsic, e.g., due to pressure inhomogeneity inside the cell, or intrinsic, due to phase separation caused by a first-order SC+CO to SC transition.

Figure 2(b) shows the pressure dependence of $T_c$  and of the zero-temperature value of $\lambda^{-2}(0)$, which is directly proportional to the superfluid density $n_s$. Interestingly, both $\lambda^{-2}(0)$ and $T_c$ show a similar non-monotonic pressure dependence, featuring three distinct regions marked as I, II and III in Fig 2(b). Initially, as pressure increases, $T_c$ shows an increase from 2.85(9)~K at ambient pressure to 6.9(3)~K at a critical pressure $p_{\rm c1}$~$\simeq$~0.63~GPa. Upon further increase in pressure, $T_c$ shows a sharp and substantial decrease down to 2.87(7)~K at 0.8~GPa. This marks the onset of Region II, where $T_c$ increases only slightly with pressure. Remarkably, a sudden jump in $T_c$ is seen around a second critical pressure $p_{\rm c2}$~$\simeq$~1.74~GPa, reaching a maximum value of $T_c^{\rm max}$~=~8.0(1)~K -- which is almost three times larger than the ambient pressure value -- and nearly saturating beyond this pressure. This regime is denoted as Region III in the $T-p$ phase diagram. 

The maximum $T_c$ value of 8~K extracted from $\mu$SR is fairly close to the value obtained through electrical transport, AC and DC magnetization measurements \cite{chen2021double}. Moreover, $\lambda^{-2}(0)$ qualitatively shows a similar trend to $T_c$ with regard to pressure, also featuring a double-peak behavior. Indeed, there is a nearly threefold increase in the value of $\lambda^{-2}(0)$ from 6.5(1)~$\mu \rm{m}^{-2}$ at ambient pressure to 17.3(9)~$\mu \rm{m}^{-2}$ at 1.74~GPa, followed by saturation upon further increase in pressure. The relative variation of the $\lambda^{-2}(0)$ value at the highest applied pressure compared to its ambient pressure value is $\delta\lambda^{-2}(0)/\lambda^{-2}(0)$~=~63~\%. This is rather unusual in comparison to BCS superconductors, where the superfluid density either depends weakly on pressure or remains pressure independent\cite{das2021unconventional}. The sudden enhancement of $T_c$ and $\lambda^{-2}(0)$ has an intimate connection with the collapse of CO at the critical pressure $p_{\rm c2}$, and is consistent with a first-order transition from the SC+CO phase to the pure SC phase.

\begin{figure*}[htb!]
\includegraphics[width=1\linewidth]{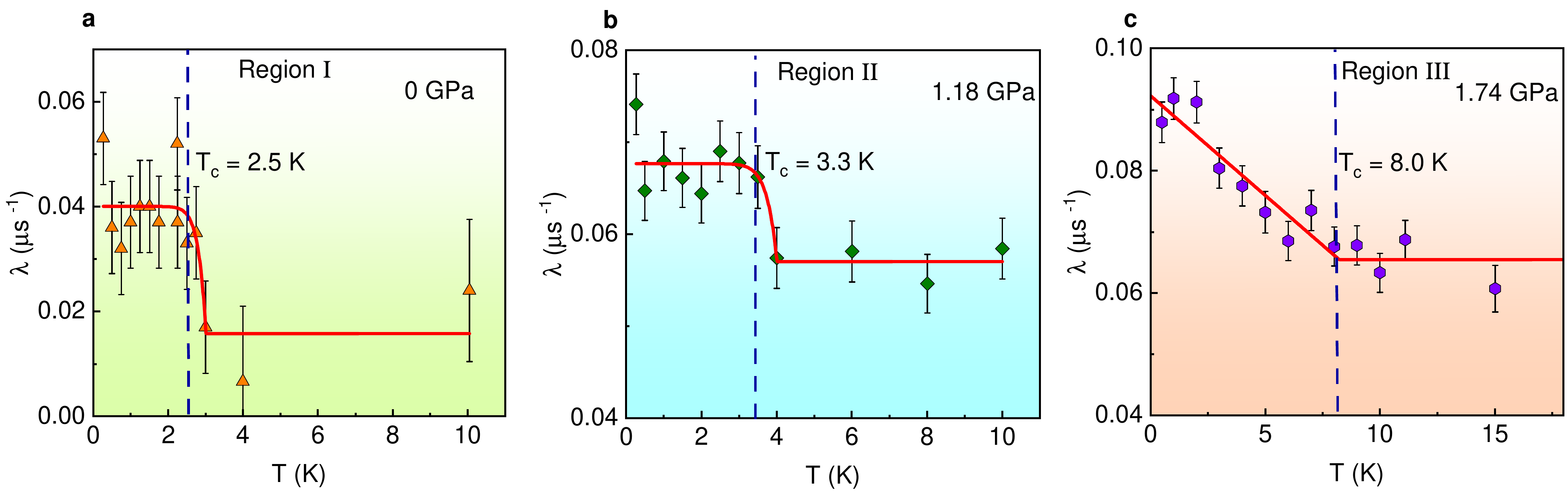}
\caption{(Color online) {\bf Indication of time-reversal symmetry-breaking in the superconducting state of CsV$_3$Sb$_5$.} a.) Electronic relaxation rate $\Lambda(T)$ for CsV$_3$Sb$_5$ as a function of temperature obtained from ZF-$\mu$SR experiments carried out in the presence of three different applied pressures: (a) 0, (b) 1.18 and (c) 1.74 Gpa. Each pressure corresponds to a different Region (I, II, and III) of the $T-p$ phase diagram of Fig.~2 (c). Solid lines through the data points are fits as described in the Methods section. The values for the superconducting transition temperature $T_c$ are denoted by the vertical dashed black line.  }
\label{fig:3}
\end{figure*}

\subsection{Superconducting transition temperature vs. superfluid density }
Figure 2(c) displays the correlation between the superconducting transition temperature $T_c$ and the superfluid density $\lambda^{-2}(0)$ obtained from $\mu$SR experiments under different pressures. There are three distinct types of $T_c (\lambda^{-2}(0))$ lines, each associated with one of the three regions of the phase diagram in Fig~2(b). For the data points belonging to Region I, we obtain a slope of 2.72~K$\mu$m$^2$, which is 3.5 times larger than the slope obtained for the data points from Region II (0.78~K$\mu$m$^2$). For Region III, $\lambda^{-2}(0)$ seems to be almost independent on $T_c$. 

These linear relationships between $T_c$ and  $\lambda^{-2}(0)$ are characteristic of unconventional superconductors. In the present case, for Regions I and II, the $T_c/\lambda^{-2}(0)$ ratio lies far away from the values typically seen in BCS superconductors, where $T_c/\lambda^{-2}~\approx$~0.00025–0.015~K$\mu$m$^2$. The large values of $T_c$, despite the small values of the superfluid density in Regions I and II, are also signatures of unconventional SC, where $T_c/\lambda^{-2}(0)\sim$~1–20 \cite{PhysRevLett.62.2317,Guguchia2022,PhysRevMaterials.5.034803}. Moreover, the change in the slope $T_c/\lambda^{-2}(0)$ from Region I to Region II, both of which display coexisting CO and SC, suggests a subtle modification in the CO state. Combined with the sudden suppression in $T_c$ and $\lambda^{-2}(0)$ upon crossing the Region I to Region II boundary in Fig.~2b, this provides strong evidence for a distinct competition between SC and CO in these two regions. Finally, the fact that $\lambda^{-2}(0)$ is nearly independent of $T_c$ in Region III can be attributed to the absence of a competing CO state in this region.

\subsection{Spontaneous fields in the superconducting state}

Zero-field $\mu$SR experiments have been successfully implemented to probe TRSB in well-known unconventional superconductors, e.g., Sr$_2$RuO$_4$\cite{luke1998time,grinenko2021split,grinenko2021unsplit}, UPt$_3$\cite{luke1993muon},  LaNiC$_2$\cite{hillier2009evidence}, LaPt$_3$P\cite{biswas2021chiral}. We have conducted ZF-$\mu$SR experiments for three different applied pressures: 0, 1.18, and 1.74~GPa. Representative ZF-$\mu$SR spectra above and below $T_c$ for the pressures of 1.18 and 1.74~GPa are shown in Figs.~1(c)~and~(d). Both display an increase in relaxation within the superconducting state. The data were fitted using the Gaussian-Kubo-Toyabe~(GKT) depolarization function \cite{Toyabe} multiplied by an exponential decay function exp$(-\Lambda t)$. The Gaussian-Kubo-Toyabe function accounts for the random magnetic fields created by the nuclear moments.  As was discussed previously for KV$_{3}$Sb$_{5}$ \cite{mielke2021time}, the exponential relaxation rate $(-\Lambda t)$ is sensitive to the temperature dependence of the electronic contribution to the muon spin relaxation. The temperature dependences of the electronic relaxation rate $\Lambda(T)$ for the applied pressures of 0, 1.18 and 1.74~GPa are shown in Fig.~3.

At ambient pressure, we observe an enhancement of the exponential relaxation rate, with an onset slightly above $T_{\rm c}$, by a small amount of $\Delta \Lambda_{\rm 0}$~=~0.024(8)~$\mu \rm{s}^{-1}$. Since time-reversal symmetry is already broken by charge order at $T_{\rm CO}$~$\simeq$~95~K \cite{yu2021evidence,RustemCVS,Wukerr}, we speculate that the effect in the superconducting state is caused by the Meissner screening of the small fields induced by the CO loop currents created at $T_{\rm CO}$.  A similar argument holds for 1.18~GPa (Region II), where both CO and SC coexist and the associated increase in electronic relaxation is $\Delta \Lambda_{\rm 1.18}~=~0.011(2)$~$\mu \rm{s}^{-1}$. 

However, at  1.74~GPa (Region III), where the CO is completely suppressed, we still observe an increase in $\Lambda(T)$ with the onset of SC at $T_{\rm c}$~$\simeq$~8~K. The increase of $\Lambda(T)$ with decreasing temperature in Region~III is smooth, contrasting with the sharp increase observed in Regions I and II. As there is no CO at this pressure, a plausible origin for the enhancement of the internal field width is the spontaneous breaking of time reversal symmetry in the superconducting state. 
The increase in $\Lambda$  by a magnitude of $\Delta \Lambda_{\rm 1.74~GPa}~=~0.031(3)~\mu \rm{s}^{-1}$ is associated with a characteristic field strength of $\Delta B$~=~$\Delta\Lambda_{\rm 1.74~GPa}$/$\gamma_\mu$~=~0.04~mT. This value is comparable to what is observed in well-known TRSB superconductors, such as 0.05~mT for the chiral superconductor Sr$_2$RuO$_4$\cite{grinenko2021unsplit,grinenko2021split}, 0.01~mT for the heavy-fermion superconductor UPt$_3$\cite{luke1993muon}, and 0.01~mT for the non-centrosymmetric superconductor LaNiC$_2$ \cite{hillier2009evidence}. These results indicate that the SC state without charge order breaks time-reversal symmetry. Interestingly, a recent $\mu$SR investigation of the other kagome compounds, RbV$_3$Sb$_5$ and KV$_3$Sb$_5$, also reported evidence for TRSB in the pure SC state \cite{Guguchia2022}.

\section{Discussion}
One of the intriguing findings of this paper is the observation of a double superconducting dome inside the region where SC and CO coexist, associated with a more than threefold enhancement of $T_{\rm c}$ and of the superfluid density, $\sigma_{\rm sc}(0)~\propto~n_s$. Indeed, Fig.~2(b) shows that with increasing pressure, $T_{\rm c}$ and the superfluid density first increase up to 0.63~GPa, followed by a sudden drop in these parameters with further increase in pressure, defining the boundary between Regions I and II. Within Region II, $T_{\rm c}$ and $n_s$ increase slowly with increasing pressure up to 1.54~GPa, beyond which a dramatic enhancement is seen in these two quantities, marking the onset of Region III. 
This behavior is suggestive of a state of coexistence between competing SC and CO orders, which undergoes a transition to a pure SC state at $p_{\rm c2}$. Such a scenario would naturally explain why not only $T_c$ but also $n_s$ displays a sudden change from Region II to Region III, as $n_s$ is suppressed in the coexisting state due to the partial gapping of the Fermi surface promoted by CO order. 

Analogously, the strong variation of $T_c$ and $n_s$ between $p_{\rm c1}$ and $p_{\rm c2}$ is indicative of a subtle modification in the nature of either the CO or the SC state from Region I to Region II. The SC state seems qualitatively the same in both regions: first, the $\lambda^{-2}(T)$ curves are indicative of a fully-gapped SC state in both regions. Second, the temperature dependence of the electronic relaxation $\Lambda(T)$ is very similar in both Regions I and II (see Figs.~3(a)~and~3(b)). Consequently, it is plausible that the changes in $T_c$ and $n_s$ are a consequence of a change in the CO state, in such a way that the CO of Region II competes more strongly with SC than the CO of Region I. Experimentally, it has been proposed that more than one CO state can be realized in these kagome metals\cite{wang2021charge,wang2021electronic,zhao2021}. Different theoretical calculations also suggest multiple nearby CO instabilities \cite{ratcliff2021coherent,Denner2021,feng2021chiral,Lin2021,christensen2021theory,Subedi2022,Balents21}. To investigate the possibility of a change in the nature of the CO state from Region I to Region II in CsV$_3$Sb$_5$, we performed DFT calculations at different hydrostatic pressures.
\begin{figure*}[htb!]
\includegraphics[width=1\linewidth]{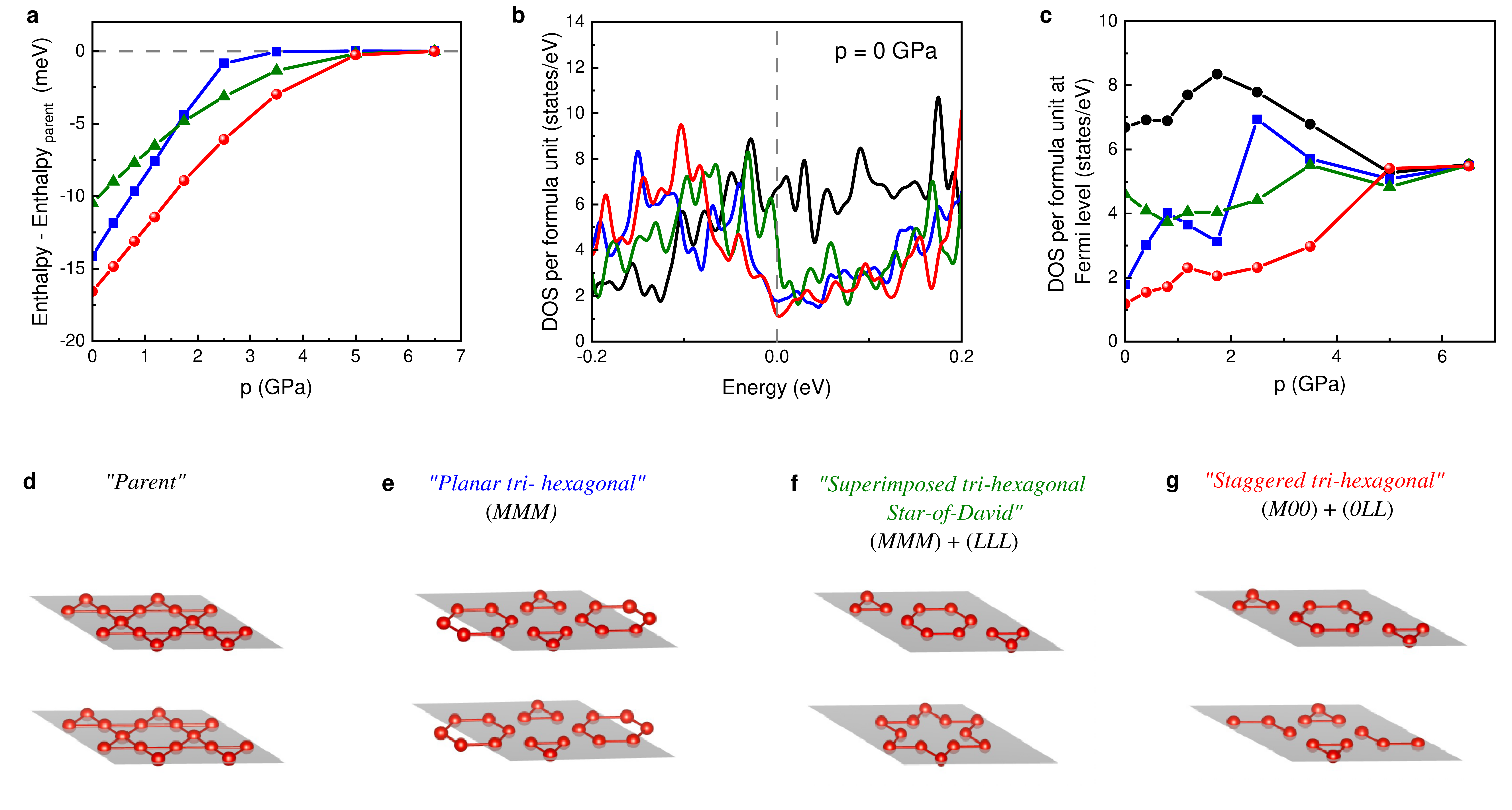}
\caption{(Color online) {\bf Theoretical modeling and DFT results.} a.) Pressure dependence of the difference in enthalpies between a charge-ordered state and that of the undistorted kagome lattice for different possible CO phases discussed in the text and shown in panels (e)-(g). b.) Density of states (DOS) per formula unit calculated at ambient pressure for the different CO phases considered. c.) Variation of the DOS per formula unit at the Fermi level with pressure. The distinct types of CO states considered here display different bond order configurations and are named:  d.) parent; e.) planar tri-hexagonal; f.) superimposed tri-hexagonal Star-of-David; and g.) staggered tri-hexagonal. Black, blue, green and red points in panels (a)-(c) correspond to these CO configurations respectively. 
}
\label{fig:4}
\end{figure*}

The $A$V$_3$Sb$_5$ compounds are predicted to have phonon instabilities at the $M$ and $L$ points of  their hexagonal Brillouin zones \cite{cho2021emergence,qian2021revealing}. Both of the unstable phonons transform as three-dimensional irreducible representations of the space group, which means that the high-temperature undistorted kagome lattice may lower its energy via lattice distortions along six independent atomic displacement patterns. Superpositions of these patterns give rise to a large family of candidate structures and bond order patterns, a subset of which are shown in Fig. 4  \cite{christensen2021theory}.

While there is growing evidence of time-reversal symmetry breaking in the CO phase, the predominant effect of pressure is expected to be on the lattice, and not directly on the properties of the CO state affected by TRSB. This is because interatomic force constants depend strongly on the bond lengths in ionic compounds, where there is a fine balance between the repulsive and the attractive components of the interatomic interactions. Moreover, the TRSB CO state is generally expected to induce a regular (i.e. time-reversal symmetry preserving) CO pattern of bond distortions due to nonlinear terms in the free energy \cite{Lin2021,Balents21}. For these reasons, despite the fact that our first-principles approach cannot directly account for TRSB, understanding how the crystal structure of CsV$_3$Sb$_5$ evolves under pressure using non-spin-polarized DFT calculations is a viable approach to elucidate the behavior of the CO in this material. 

In Fig.~4(a), we present the enthalpy per formula unit of the three lowest energy CO structures of CsV$_3$Sb$_5$ calculated from DFT as a function of pressure. The parent structure, without CO, is shown in Fig.~4(d). Two of these CO phases, dubbed the $(M00)+(0LL)$ ``staggered tri-hexagonal'' state (red, pane~ (g)) and the $(MMM)+(LLL)$ ``superimposed tri-hexagonal Star-of-David'' phase (green, panel~(f)), host a superposition of both $M$ and $L$ point lattice instabilities, and therefore have the $2\times2\times2$ periodicity consistent with the available X-ray data. The third low energy phase, the so-called $(MMM)$ ``planar tri-hexagonal" phase (blue, panel~(e)), is only associated with the $M$ point instability. As a result, it has a $2\times 2\times 1$ periodicity that seems inconsistent with the currently available experimental structure data. 

Fig.~4(a) shows that at pressures above 5~GPa, CO is completely suppressed, and we predict the undistorted kagome lattice to be the equilibrium structure. This is consistent with previous results that predicted the phonon instabilities at both $M$ and $L$ to vanish at pressures between 3~GPa and 5~GPa \cite{zhang2021first}. The threefold difference with respect to the experimental $p_{c,2}$ value is not unexpected. Indeed, an accurate prediction of the pressure at which the structural instabilities disappear is hard to achieve in DFT because of the large unit cell, the sensitivity of the electronic structure of the kagome layers on pressure, and lattice parameters errors due to the exchange correlation approximations. Nevertheless, the qualitative trends with pressure extracted from DFT are expected to be reliable.

While the staggered tri-hexagonal phase (red) has lower enthalpy than the other CO phases throughout the pressure range considered, the enthalpy differences between the distinct CO phases are on the order of a few meV's per formula unit. This implies that entropic effects not captured by DFT are large enough to switch the ordering of these phases and cause the equilibrium structure to be a different phase from the one that minimizes the Kohn-Sham enthalpy. The two most likely candidates consistent with the experimentally observed $2\times~2~\times 2$ periodicity of the CO state are the staggered tri-hexagonal (red) and the superimposed tri-hexagonal Star-of-David phases (green). Since the energy difference between these phases become smaller with increasing pressure, it is plausible that there is a transition from one to the other under pressure.

To elucidate which of these two CO phases is realized in Region I, we compute their densities of states (DOS). As shown in Fig.~4(b), there are important differences between the DOS curves of the staggered tri-hexagonal phase (red), the superimposed tri-hexagonal Star-of-David phase (green), and the undistorted kagome lattice (black). In particular, the latter has a larger DOS at the Fermi level, as well as van Hove singularities (signaled by the central ``satellite" peaks in the DOS) closer to the Fermi level, when compared to the red and green curves (the blue curve, referring to the planar tri-hexagonal phase, is also shown for completeness). In contrast, the red curve has the smallest Fermi level DOS and farthest separated van Hove singularity peaks among the three curves. While the mechanism for superconductivity in CsV$_3$Sb$_5$ remains under debate, one generally expects that a higher Fermi-level DOS correlates with a higher $T_c$. Moreover, if van Hove singularities are important for the SC mechanism, it is reasonable to assume that the closer their peaks are to the Fermi level, the higher the $T_c$ will be. If these assumptions hold for CsV$_3$Sb$_5$, then a possible explanation for the unusual double-peak SC dome revealed by our $\mu$SR data is a transition from the superimposed tri-hexagonal Star-of-David phase (green) in Region I to the staggered tri-hexagonal phase (red) in Region II. This is because the latter should compete more strongly with SC due to the larger reduction in the DOS and further separation between the van Hove peaks. Similarly, because the undistorted lattice (i.e. without CO) has the highest DOS, this would also account for $T_c$ being the highest when all types of CO are suppressed at $p_{c,2}$.

\section{Conclusion} 

  Our results uncover two different regions in the part of the phase diagram of the kagome metal CsV$_3$Sb$_5$ where both charge order and superconductivity are present. Based on our DFT calculations, we propose that two distinct types of CO order are realized: a superimposed tri-hexagonal Star-of-David phase at low pressures and a staggered tri-hexagonal phase at intermediate pressures, before CO is fully suppressed. 
These charge orders display different degrees of competition with superconductivity, which leads to the complex phase diagram featuring a double-peak in the pressure dependence of both the superconducting transition temperature and the superfluid density. Throughout the pressure range investigated, the nodeless multigap nature of the superconducting state remains persistent, in contrast to KV$_3$Sb$_5$ and RbV$_3$Sb$_5$, where a nodal gap emerges at higher pressures \cite{Guguchia2022}. Furthermore, we show that once charge order is fully suppressed, the superconducting state breaks time-reversal symmetry, which makes this compound, together with its K- and Rb-counterparts, one of the rare cases of unconventional superconductors with spontaneously broken time reversal symmetry.

\section{Methods} 

\textbf{Sample growth:} Single crystals of CsV$_3$Sb$_5$ used for for this study were grown by using self-flux method from Cs ingots (purity 99.9\%), V 3-N powder (purity 99.9\%) and Sb grains (purity 99.999\%)\cite{yin2021superconductivity}. The detailed description about the growth and characterization of crystals is provided in detail in Refs. \cite{gupta2021microscopic,ortiz2020cs,ortiz2019new}

\textbf{$\mu$SR experiment:} Muon spin relaxation/rotation ($\mu$SR) measurements under pressure were performed at $\mu$E1 beam line using GPD spectrometer at Paul Scherrer Institute, Switzerland. This spectrometer is a dedicated worldwide unique instrument for carrying out high pressure $\mu$SR experiments. For $\mu$SR experiments, 100~\% spin polarized muons $\mu^+$ are implanted inside the sample one by one, where the spin of muon will Larmor precess around the local magnetic field $B_{\rm int}$ at the muon site. Detailed description about the $\mu$SR technique can be found in the Refs.\cite{yaouanc2011muon,HillierMSR} In case of a superconductor, the characteristic length scale namely penetration depth can be determined by probing the inhomogeneous field distribution of vortex lattice. Heliox cryostat with a He-3 insert was used to access temperature down to 250~mK. Pressure upto 1.9~GPa were achieved using low and pre-defined background double wall pressure cell made from MP35N/CuBe alloy, specifically designed for high pressure experiments \cite{shermadini2017low,khasanov2016high}. Single crystals of CsV$_3$Sb$_5$ were filled with random orientations inside the pressure cell in a compact cylindrical area of height 12~mm and diameter 6~mm. Daphne 7373 was used as a pressure medium to ensure the hydrostatic pressure conditions as it solidifies at much higher pressure $\sim$ 2.5~GPa. Low temperature values of pressures were determined by tracking the superconducting transition of Indium, measured using ac susceptibility experiment. The fraction of muon stopping inside the sample was maximized to $\sim$ 40~\%.

  We performed transverse-field (TF) and zero-field ${\mu}$SR experiments at different pressures. Pressure cell is surrounded by four detectors: Forward, backward, left, and right with muon initial spin parallel to its momentum. The TF-${\mu}$SR experiments were performed in field cooled condition to get homogeneous vortex lattice formation. Approximately 10$^6$ positrons were recorded for each data point. The asymmetry vs. time spectra were analyzed using the open software package \textsc{Musrfit} \cite{suter2012musrfit}.

 \textbf{Analysis of $\lambda(T)$:} The temperature variation of magnetic penetration depth $\lambda^{-2}(T)$ is analyzed within the local (London) approach using following expression\cite{serventi2004effect}:
\begin{multline}
\frac{\lambda^{-2}(T,\Delta_0)}{\lambda^{-2}(0,\Delta_0)}=1+\frac{1}{\pi}\int_{0}^{2\pi}\!\!\int_{\Delta (T, \phi)}^{\infty}\frac{\partial f}{\partial E } \frac{E\,\mathrm{d}E\,\mathrm{d}\phi}{\sqrt{E^2-\Delta (T, \phi)^2}},
\end{multline}
where $f=(1+E/k_\mathrm{B}T)^{-1}$ is the Fermi distribution function. The temperature and angular dependence of gap is given by  $\Delta (T, \phi)=\Delta_0\delta(T/T_c)g(\phi)$, where $\Delta_0$ is the gap value at 0 K, $\delta(T/T_c)=\tanh\{1.821[1.018(T_c/T-1)^{0.51}]\}$ represents the temperature dependence of gap. The angular dependence of the gap is represented by $g(\phi)$, which adapts a value 1 for isotropic $s$- wave model. We have fitted the $\lambda^{-2}(T)$ data at various pressures by separating it into two gaps:
\begin{equation}
\frac{\lambda\textsuperscript{-2}(T)}{\lambda\textsuperscript{-2}(0)}= x\frac{\lambda\textsuperscript{-2}(T,\Delta_{0,1})}{\lambda\textsuperscript{-2}(0,\Delta_{0,1})}+(1-x)\frac{\lambda\textsuperscript{-2}(T,\Delta_{0,2})}{\lambda\textsuperscript{-2}(0,\Delta_{0,2})}.
\end{equation}
In the above equation, $x$ is the weight factor of the bigger gap $\Delta_{0,1}$. Fig.~S3 shows the pressure dependence of the superconducting gaps ($\Delta_{0,1}$ and $\Delta_{0,2}$). In the regions where SC and CO coexists, i.e. Region I and II, fitting was done by considering the weight factor $x$ as a global parameter for all pressures in the mentioned regimes. Similarly, for Region III, the fitting was done by assuming $x$ as a global parameter in that region. The data at intermediate pressure 1.68~GPa is fitted by considering a linear combination of Eq.~2 with parameter values fixed from that obtained for 1.54~GPa (corresponding to the SC+CO state) and 1.74~GPa (corresponding to the SC state).

\textbf{Analysis of ZF relaxation rate:} $\Lambda(T)$ at various pressures are fitted with following empirical relation:
\begin{equation}
   \lambda(T) =
\begin{cases}
\Lambda_0  , & T>T^* \\
\Lambda_0+\Delta\Lambda\left[1-\left(\frac{T}{T^*}\right)^n\right] , & T<T^*
\end{cases}.
 \label{eq:lambda}
\end{equation}
where $T^*$ is a characteristic temperature below which the increase in electronic  relaxation starts. $\Lambda_0$ is the relaxation value above $T_c$, and $\Delta\Lambda$ is the change in relaxation after entering into the superconducting ground state. In Region I and II, the value of $n$ is considerably high (14 and 15 respectively) indicating the step like increase in the relaxation rate. On the other hand, in Region III, where CO suppress completely, the value of $n$ is 1 suggesting a linear increase in $ \lambda(T)$ below $T^*$ which coincides with $T_c$. As discussed in the main text, we conclude that in region III, $T^*$ corresponds to the time reversal symmetry breaking transition $T_{\rm TRSB}$. 

\textbf{DFT-calculations:} All DFT calculations were performed using Projector Augmented Waves (PAW) as implemented in the Vienna Ab initio simulation package (VASP) version 5.4.4 \cite{kresse1993ab,kresse1996efficiency,kresse1996efficient}. We used the PBEsol exchange correlation functional, with valence configurations of 5\emph{s}$^2$5\emph{p}$^6$6\emph{s}$^1$, 3\emph{s}$^2$3\emph{p}$^6$3\emph{d}$^4$4\emph{s}$^1$, and 5\emph{s}$^2$5\emph{p}$^3$ corresponding to Cs, V, ans Sb, respectively. Lattice parameters were found to be converged to within 0.001 {\AA} using a plane wave cutoff energy of 450 eV, a Monkhorst-Pack k-point mesh of 20$\times$20$\times$10 in the primitive cell, and a 2$^{nd}$ order Methfessel-Paxton smearing parameter of 10 meV \cite{methfessel1989high}. Different phases have different specific volumes, and hence at finite pressure, the phase that minimizes the \emph{enthalpy}, rather than the Kohn-Sham energy only, is the equilibrium structure at that volume. 

Bilbao Crystallographic Server and tools therein were used for symmetry analysis of various theoretically obtained crystal structures \cite{aroyo2011}.

{\bf Acknowledgments}\\
We thank B. Andersen and M. Christensen for fruitful discussions. ${\mu}$SR experiments were performed at the Swiss Muon Source (S$\mu$S), Paul Scherrer Institute (PSI), Switzerland. H.C.L. was supported by National Key R\&D Program of China (Grant No.\ 2018YFE0202600) and the Beijing Natural Science Foundation (Grant No.\ Z200005). T.B. and E.R. were supported by the NSF CAREER grant DMR-2046020. RMF (theory) was supported by the Air Force Office of Scientific Research under award
number FA9550-21-1-0423.

{\bf Author contributions}\\
R.G., D.D., C.M., Z.G., F. H., and R.K. performed ${\mu}$SR experiments. R.G., D.D. and R.K. analyzed the ${\mu}$SR data. Q.Y., Z.T., C.G., and H.C.L. synthesized and characterized samples. R. M. F., T. B., and E.R. performed the theoretical analysis. R. K. and H.L. supervised the work at PSI. R.G., D.D., Z.G., and R. M. F. prepared the manuscript with notable inputs from all authors.

{\bf Competing interests}\\
The authors declare no competing interests.

{\bf Data availability:}\\
The data supporting the findings of this study are available within the paper and in the Supplementary Information. The raw data are available from the corresponding authors upon reasonable request.


\end{document}